\documentclass[preprint,showpacs,preprintnumbers,amsmath,amssymb]{revtex4}

\usepackage{graphicx}
\usepackage{dcolumn}
\usepackage{bm}

\begin{document}


\title{Nanohertz Frequency Determination\\for the Gravity Probe B HF SQUID Signal}

\author{M. Salomon}
\email{michael.salomon@stanfordalumni.org}
\affiliation{Department of Aeronautics \& Astronautics, Stanford University,
Durand Building, 496 Lomita Mall, Stanford, CA  94305-4035\\
}%

\author{J.W. Conklin}%
\email{johnwc@stanford.edu}
\affiliation{Department of Aeronautics \& Astronautics, Stanford University,
Durand Building, 496 Lomita Mall, Stanford, CA  94305-4035\\
}%

\author {J. Kozaczuk}
\email{jkozaczu@ucsc.edu}
\affiliation {Department of Physics, University of California Santa Cruz,
211 Interdisciplinary Sciences Building,
1156 High Street Santa Cruz, CA 95064}

\author{J.E. Berberian}
\email{berberian@alum.mit.edu}
\affiliation{
Berberian \& Company, LLC, 3865 Wilson Blvd Ste 520, Arlington, VA 22203-1764\\
}%

\author{G.M. Keiser}
\email{mackeiser@gmail.com}
\affiliation{Hansen Experimental Physics Laboratory, Stanford University,
452 Lomita Mall, Stanford, CA 94305-4085\\
}%

\author{D.I. Santiago}
\email{davitivan@gmail.com}
\affiliation {
Berberian \& Company, LLC, 3865 Wilson Blvd Ste 520, Arlington, VA 22203-1764\\
}%

\author{A.S. Silbergleit}
\email{gleit@stanford.edu}
\affiliation{Hansen Experimental Physics Laboratory, Stanford University,
452 Lomita Mall, Stanford, CA 94305-4085\\
}%

\author{P. Worden}
\email{pworden@stanford.edu}
\affiliation{Hansen Experimental Physics Laboratory, Stanford University,
452 Lomita Mall, Stanford, CA 94305-4085\\
}%

\date{\today}

\def \Id{$C_1$ }
\def \Idd{$C_2$ }
\def \Iddd {$\left(C_1,C_2\right)$ }
\begin{abstract}
In this paper, we present a method to measure the frequency and the
frequency change rate of a digital signal.
This method consists of three consecutive algorithms:
frequency interpolation, phase differencing,
and a third algorithm specifically designed and tested by the authors.
The succession of these three algorithms allowed a 5 parts in $10^{10}$
resolution in frequency determination. The algorithm developed by the authors
can be applied to a sampled scalar signal such that a model linking the
harmonics of its main frequency to the underlying physical phenomenon
is available.
This method was developed in the framework of the Gravity Probe B (GP-B)
mission. It was applied to the High Frequency (HF) component of GP-B's
Superconducting QUantum Interference Device (SQUID) signal,
whose main frequency $f_z$ is close to the spin frequency of the gyroscopes
used in the experiment. A $30$ nHz resolution in signal frequency and a
$0.1$ pHz/sec resolution in its decay rate were achieved out of a
succession of $1.86$ second-long stretches of signal sampled at $2200$ Hz.
This paper describes the underlying theory of the frequency measurement method
as well as its application to GP-B 's HF science signal.
\end{abstract}



\maketitle
\section{Background and available signal}
\label{part1:spin_computation}

The GP-B experiment aims at testing in Earth orbit two predictions of
Einstein's general relativity using precision gyroscopes.
This idea was independently proposed by Pugh \cite{pugh} and
Schiff \cite{schiff1} \cite{schiff2} in 1960.
Both of these authors pointed out that according to the general
theory of relativity, the angular momentum axis of a gyroscope
in orbit about the Earth will precess about a direction normal
to the orbital plane due to the gravitational interaction of the spinning
gyroscope with its orbital motion, and simultaneously about the direction
of the Earth's rotation axis due to the interaction of the spinning
gyroscope with the angular momentum of the Earth.
The first effect is known as the geodetic effect,
and the second is known as the frame-dragging effect.
On a 640 km polar orbit, the gyroscope drift rate due to the orbital 
motion about the Earth is 6.6 arcsec/yr (32 $\mu$rad/yr),
while the orbital average drift rate due to the
Earth's angular momentum is 0.041 arcsec/yr (0.20 $\mu$rad/yr). 

GP-B uses four gyroscopes spinning in a quasi torque-free environment
and placed inside a drag-free satellite. The orientation of the
gyroscopes is known thanks to an on-board telescope pointing towards a
distant guide star, whose orientation with respect to an extra-galactic
source is known.  Furthermore, the satellite rolls about the telescope axis.
As the orientation of the satellite with respect to an inertial reference
frame is nonetheless known, reference frames linked to the satellite are
qualified as inertial throughout this paper.
GP-B's scientific goal can be fulfilled by measuring the
orientation of the gyroscope's spin axis with respect to the satellite. 

The gyroscopes are superconductive, which allows tracking of the orientation
of their angular velocity vectors. Indeed, a spinning, superconducting
body creates a magnetic dipole parallel to its spin axis,
the London moment \cite{becker} \cite{lond}.
This magnetic dipole is then an excellent
indicator of the direction of the instantaneous spin axis.
Since the gyroscopes are almost perfectly spherical and uniform
($\Delta I \le 10^{-6}$),
the spin axis direction is a very good indicator of the direction of the
angular momentum.

Low-noise Superconducting QUantum Interference Device (SQUID) magnetometers
are thus used to measure the magnetic flux
through a pick-up loop placed around each gyroscope
created by the London moment plus a contribution due
to a magnetic field trapped in the rotor.
The SQUID signal is proportional to the magnetic flux through the pick-up loop.
The SQUID output is an analog signal which, on board the satellite,
is split into a low frequency (LF) channel---which contains the London moment
contribution---and a high frequency (HF) channel.
Both HF and LF channels pass through a $780$ Hz low pass analog filter.
The LF channels then passes through an additional 4 Hz analog low pass filter
and an additional gain stage.
In this paper, we are only concerned about the HF channel,
which is sampled at $2200$ Hz and digitized with a
16 bit ADC with a range of $\pm$~10~V.
This ADC provides a resolution higher than the signal-to-noise ratio which
is $\sim \: 10^5$ near the gyroscope spin frequency.
We refer to this digitized
high resolution signal as the "HF SQUID signal".

When each gyroscope transitioned below its critical temperature,
the flux due to the residual magnetic field surrounding it before
the transition was trapped on its surface, forming a large number
of small magnetic sources. These sources are called fluxons,
and can be pictured as rigidly linked to the surface of the body,
as shown for instance in \cite{cabr}.
A consequence is that the fluxons exactly follow the motion of
the body and create modulations, on the order of a few volts,
at a frequency close to the
spin frequency of the gyroscope: these modulations constitute the HF signal.

The HF SQUID signal is sent intermittently in the form of 1.86 second
long stretches that we refer to as "snapshots".
Each snapshot contains 4096 points.
We also have access to this signal in the form of
Fast Fourier Transforms (FFTs) performed at regular intervals on the
SQUID signal by the on-board CPU.
The FFT is applied every 10 seconds to sets of 4096 points of raw data,
and a compacted form of its output is sent to the ground. 


An on-board FFT algorithm is applied to $1.86$ second-long stretches
of HF SQUID signal and can thus resolve frequencies $1/1.86$ sec =$0.54$ Hz
apart. This frequency resolution, called a 'bin', is poor:
the FFT data is thus further processed.
This additional processing is performed on the ground,
and requires computing the FFT at a few frequencies.
The value of the FFT at the first five harmonics of the signal frequency
and the one corresponding to the $110$ Hz calibration signal
are thus telemetered. In addition, the value of the FFT at the two bins
adjacent to each one of these six frequencies are also sent down,
as well as the zero-frequency term.
For each $1.86$ second-long data stretch,
the on-board FFT algorithm thus provides 19 data points.
Since the high frequency data is sampled at $2200$ Hz,
the initial data stretch contains $4096$ points and so does its FFT,
so transmitting 19 data points is a dramatic reduction in bandwidth. 

\begin{figure}
  \begin{center}
    \includegraphics[width=10cm]{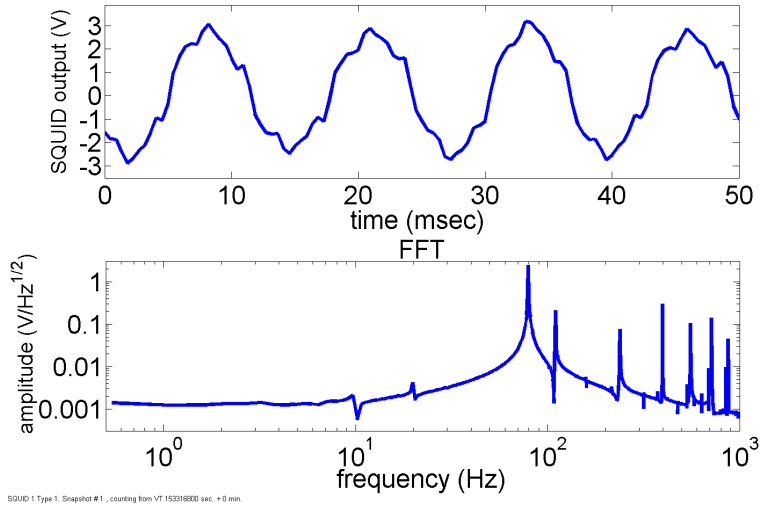}
  \end{center}
  \caption{\label{fig:snapshot}(Color online)
    Snapshot (time-series and spectrum) for Gyroscope 1,
    taken November 10, 2004 }
\end{figure}

Additionally, raw snapshots of the SQUID signal are also sent down on an
irregular but frequent basis. These snapshots also consist of
$1.86$ second-long data stretches sampled at $2200$ Hz.
An example of the time history of a portion a snapshot is provided in the
upper pane Figure \ref{fig:snapshot},
while the FFT of the entire shapshot is shown in the lower pane.

In this paper, we describe the method that was used to determine
the frequency of this HF signal, and therefore the rotor spin speed,
with a resolution better than $30$ nHz.
This was critical for the GP-B data analysis because it allowed
the time-varying readout scale factor to be determined to 1 part in $10^4$,
the orientation of the spin axis with respect to the spacecraft to be
determined to $\sim$ 3 marcsec in 1 orbit,
and the gyroscopes' relativistic drift rates to be determined
to $20$ marcsec/yr \cite{everitt2011}.
The time-varying scale factor is caused by the trapped flux contribution to
the magnetic flux through the pick-up loop, which varies at
the rotor spin $\pm$ spacecraft roll frequency, the rotor polhode frequency
and at low frequency.
With an accurate estimate of the rotor polhode and the estimate of the
rotor spin speed to 30 nHz (discussed here)
the body-fixed orientation of the gyroscope
rotor with respect to the spacecraft was determined
with an accuracy of $\sim$ 1 deg throughout the
entire science mission, lasting 1 year.
This information was necessary to determine the distribution
of trapped flux on the surface of the rotor and the time varying
readout scale factor \cite{silb}.

Before delving into the 3 successive frequency estimation algorithms,
we introduce notations specific to the GP-B experiment in order to
explain the relationship between the gyroscope motion and the SQUID signal.
We then describe how frequency interpolation and phase differencing,
a time-domain technique,
were applied to the FFT data to achieve a 5 $\mu$Hz frequency resolution.
We then show that this result was checked using the snapshot data.
Variations of these first two techniques are known
and their accuracy in determining monotone signals in the presence
of Gaussian white noise and simple systematic effects have been studied
\cite{schoukens}.
Finally, we show how the snapshot data and the result of the phase
differencing were used to run the algorithm developed by the authors which
allowed a 5 parts in $10^{10}$ accuracy in frequency determination.

\section {Gyroscope motion and HF SQUID signal frequency}

As the fluxons attached to the gyroscope move with it,
HF modulations are created in the SQUID signal.
The fluxons create a body-fixed distribution of potential:
a model for this distribution can thus be written in body-fixed frame,
for instance using a spherical harmonics expansion \cite{salo,firstTFM}.
The SQUID measurement however takes place in the satellite frame - which,
as explained above, is considered inertial.
Therefore, if the appropriate set of Euler rotations is applied to rotate
the body-fixed frame into the inertial frame, it is possible to express
the HF SQUID signal as a function of the coefficients of the model of
the magnetic potential distribution.

\begin{figure}
  \begin{center}
    \includegraphics[width = 9 cm]{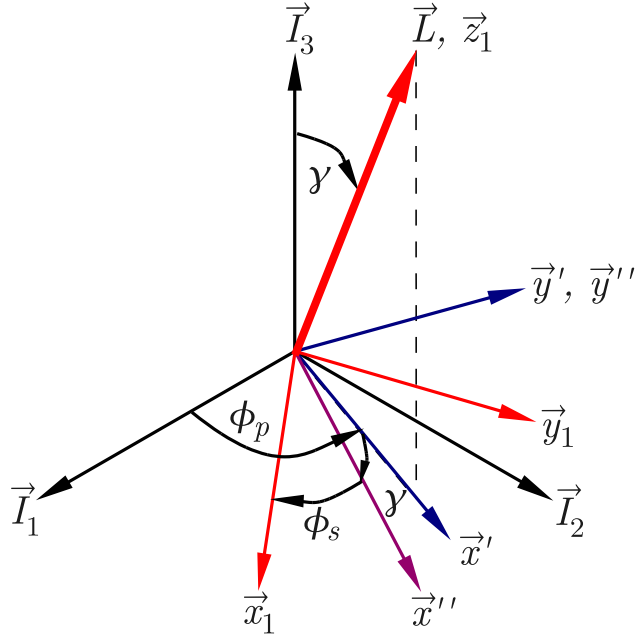}
  \end{center}
  \caption{\label{fig:BFIF_part3}(Color online)
    Rotations from body fixed frame $(\vec{I}_1,\vec{I}_2,\vec{I}_3)$
    to inertial frame $(\vec{x}_\text{I},\vec{y}_\text{I},\vec{z}_\text{I})$ }
\end{figure}

The vectors $\vec{I}_1$, $\vec{I}_2$ and $\vec{I}_3$ are the principal
inertia axes of the gyroscope and define an orthonormal
body-fixed reference frame.
We call $(\vec{x}_\text{I},\vec{y}_\text{I},\vec{z}_\text{I})$
an orthonormal inertial reference frame such that $\vec{z}_\text{I}$
is aligned with the angular momentum $\vec{L}$ of the gyroscope.
The angular momentum is inertially fixed as the gyroscopes are in
torque-free motion.
These notations are shown in figure \ref{fig:BFIF_part3}.

The first Euler rotation from $(\vec{I}_1,\vec{I}_2,\vec{I}_3)$ to
$(\vec{x}_\text{I},\vec{y}_\text{I},\vec{z}_\text{I})$
is an azimuthal rotation by an angle $\phi_p$.
This rotation transforms $(\vec{I}_1,\vec{I}_2,\vec{I}_3)$
into a reference frame $\vec{x'}\vec{y'}\vec{z'}$ whose $\vec{z'}$
axis is aligned with $\vec{I}_3$.
A polar rotation by an angle $\gamma$ is then applied:
the new reference frame $\vec{x''}\vec{y''}\vec{z''}$ is such that its
third axis $\vec{z''}$ is aligned with the angular momentum.
The third Euler angle $\phi_s$ measures the angle by which the
gyroscope has spun about $\vec{L}$ since a fixed time origin:
a rotation of angle $-\phi_s$ about $\vec{L}$ is thus also needed to
obtain the inertial reference frame
$(\vec{x}_\text{I},\vec{y}_\text{I},\vec{z}_\text{I})$. 

Consequently, the frequency at which any fluxon passes through the pick up
loop is $\dot{\phi}_s/2\pi+f_p$, where $f_p$ is the polhode frequency,
such that $f_p \sim \dot{\phi}_p/2\pi$.
As the modulations in the HF SQUID signal are due to the motion of the
fluxons with respect to the pick-up loop,
the frequency $f_z$ of this signal is thus given by:

\begin{equation}
  \label{eqn:fz}
  f_z=\frac{1}{2\pi}\dot{\phi}_s + f_p.
\end{equation}

An expression for $\dot{\phi}_s$ can be obtained from \cite{mcmi},
formulas 89.4 and 90.4:

\begin{equation}
  \label{eqn:first_dotphis}
  \dot{\phi}_s=\frac{L}{I_3} \left[ 1 + \frac{I_3-I_1}{I_1}
    \frac{1}{1+\alpha^2 sn^2(\tau,k^2)}\right],
\end{equation}
where the characteristic, $\alpha$, and the elliptic modulus, $k$,
only depend on the moments of inertia
$I_1$, $I_2$ and $I_3$, the rescaled time,$\tau$,
depends on the three moments of inertia
and the angular momentum, $L$ (see \cite{salo}) and $sn$ is the
Jacobi elliptic integral referred to as "sinus amplitude" and defined by:
$$
sn(\tau,k)=x~~\text{when}~~ \tau=\int_0^x\frac{ds}{\sqrt{(1-s^2)(1-k^2s^2)}}.
$$
The first term in Eq. (\ref{eqn:first_dotphis}) is the larger by a factor
of $10^6$, and the second term is
a modulation at twice the polhode frequency.
The three Euler rotations are summarized in the definition of the
spin axis $\vec{\omega}$:
$$
\vec{\omega}=\dot{\phi}_p\vec{I}_3 + \dot{\gamma}\vec{y'}
- \dot{\phi}_s\vec{z''}.
$$
The spin frequency $f_s$ is thus

\begin{equation}
  \label{eqn:omeg_angles}
  (2\pi f_s)^2=\dot{\phi}_p^2+\dot{\phi}_s^2+\dot{\gamma}^2 +
  2\dot{\phi}_s\dot{\phi}_p \cos{\gamma}.
\end{equation}

As $\dot{\phi}_p$ and $\dot{\gamma}$ can be as large as $f_p$ which
is on the order of $0.1$ mHz on all gyroscopes \cite{salo},
$\dot{\phi}_s$ lies within $0.1$ mHz of the spin frequency $f_s$ and
is thus on the order of 100 Hz. The HF SQUID signal frequency $f_z$
is thus close to the spin frequency $f_s$.

The first two steps of the frequency determination method presented
in this paper are applied to the HF SQUID signal in order to determine $f_z$.
The third step of the frequency determination method presented in this
paper thus aims at determining $\dot{\phi}_s$ with a $20$ nHz resolution.
In \cite{salo}, a procedure is given to measure the polhode frequency
$f_p$ with a $10$ nHz accuracy.
Note that this represents an accuracy of 1 part in $10^4$ as the polhode
frequency is on the order of 0.1 mHz. Therefore, from (\ref{eqn:fz}),
an absolute accuracy better than 30 nHz in the the
determination of the frequency $f_z$ of the HF SQUID signal can be achieved.
In relative terms, the claimed accuracy of the frequency $f_z$
determination is thus on the order of 5 part in $10^{10}$.

\section {Millihertz level signal frequency determination through
frequency interpolation}
\label{part:gnd_det_spin_speed}

By interpolating the FFT data, it is possible to go beyond the $0.54$ Hz
frequency resolution and to determine the frequency with a $1$ mHz precision.
This method is well known and has been used for several decades \cite{rife}.
It consists of approximating the signal by a pure sine around
its main frequency, and finding the frequency of such a sine if it
had the same amplitude diagram as the FFT of the signal.
We now give a mathematical description of this method.

As the interpolation is applied to the FFT data which is itself computed
using short stretches of SQUID signal, we neglect the polhode harmonics
in this description. We show, without loss of generality,
how frequency interpolation is implemented for a single frequency signal.

Let $z_{\text{HF}}(t)$ be our signal, so that
$$
z_{\text{HF}}(t)=A_s \cos{(2\pi f_z t + \delta\phi(t))}.
$$
Its discrete Fourier transform is: 
\begin{eqnarray}
  F(f)= &&\frac{A_s}{2} e^{i\delta\phi}e^{-i2\pi(f - f_z)\Delta t
    \frac{N - 1}{2}}
    \left[ \frac{\sin[ 2\pi(f - f_z)\Delta t N / 2]}
    {\sin[ 2\pi(f - f_z)\Delta t / 2]}\right] \nonumber\\
  && +\frac{A_s}{2} e^{-i\delta\phi} e^{-i 2\pi(f + f_z)
    \Delta t \frac{N - 1}{2}}
    \left[ \frac{\sin\left[ 2\pi(f + f_z)\Delta t N / 
    2\right]}{\sin\left[ 2\pi(f + f_z)\Delta t / 2\right]}\right] 
  \label{eqn:full_Fourier_Cos}
\end{eqnarray}

where:
$\Delta t$ is the time between two consecutive samples
($1/2200$ sec for a $2200$ Hz sampling rate),
$N=4096$ is the number of points in the FFT,
N$\Delta t$ is then the total snapshot duration,
$f_d = 1/(N \Delta t)$ is the bin frequency,
$f$ is the frequency at which the FFT is computed,
$A_s$ is the signal amplitude and $\delta\phi$ is the phase shift.
 
Near the HF signal's frequency, $f \approx f_z$, so the first term in
(\ref{eqn:full_Fourier_Cos}) is dominant and we neglect the second one.
Since the FFT is discrete, we only obtain its values at multiples of the
bin frequency $f_d$. Furthermore, as mentioned above,
we have at our disposal the value of the FFT at the central bin and at the
two adjacent bins. Let's define the integer $n$ such that
$nf_d$ is the multiple of $f_d$ closest to the signal frequency $f_z$. We note:

$$
F_n = F(nf_d) ~~~F_{n+1} = F[(n+1)f_d] ~~~ F_{n-1} = F[(n-1)f_d] .
$$
Then the quantity we use in the interpolation is:
$$
F_n =\frac{A_s}{2} e^{i \delta \phi} e^{-i2\pi(n f_d -f_z)\Delta t \frac{N-1}{2}}
\left[ \frac {\sin\left[2\pi(nf_d-f_z)\Delta tN/2\right]}
{\sin\left[2\pi(nf_d-f_z)\Delta t/2\right]}\right].
$$
$F_{n-1}$ and $F_{n+1}$ are defined similarly. Let's also introduce the ratios:
$$
R_{n+1}= \Big| \frac{F_{n+1}}{F_n}\Big| ~~~R_{n-1}=
\Big| \frac{F_{n-1}}{F_n}\Big|.
$$
The values of those two ratios are obtained from measurements.
We define the quantity $x_n$ to be,
\begin{equation}
  \label{eqn:x_n}
  x_n= 2\pi(nf_d-f_z)N\Delta t/2.
\end{equation}
Then, by performing a Taylor series expansion of $F_m$ in the quantity
$x_m/N$, where $m$ takes values $n-1$, $n$, $n+1$, we obtain,
\begin{equation}
\label{equ:Rn}
  R_{n+1} = \left| \frac{x_n}{x_n+\pi} \right|
    + \mathcal{O} \left(\frac{x_n}{N}\right)
    ~~~R_{n-1} = \left| \frac{x_n}{x_n-\pi} \right|
    + \mathcal{O} \left(\frac{x_n}{N}\right) .
\end{equation}

As implied by Eq. \ref{eqn:x_n},
the term $(x_n/N)$ here is $10^{-3}$ at most, since near the main peak
the signal frequency, $f_z$, is at worst one frequency bin away from $nf_d$.
From these relations, we derive two formulas for the frequency
$f_z$ of the HF SQUID signal, depending on the sign of $x_n$.
For $x_n > 0$:

\begin{equation}
f_z=nf_d +\frac{1}{N\Delta t}\frac{R_{n+1}}{R_{n+1} - 1} + \mathcal{O}
\left(10^{-3}\right),
\end{equation}
\begin{equation}
f_z=nf_d -\frac{1}{N\Delta t}\frac{R_{n-1}}{R_{n-1} + 1}
+ \mathcal{O} \left(10^{-3}\right).
\end{equation}
Averaging the two formulas for $f_z$ we and obtain:
\begin{equation}
  \label{eqn:omeg_interp}
  f_z=nf_d +\frac{1}{2N\Delta t} \left[ \frac{R_{n+1}}{R_{n+1} - 1}
  - \frac{R_{n-1}}{R_{n-1} + 1}\right] +\mathcal{O} \left(10^{-3}\right)
\end{equation}
For $x_n < 0$:

\begin{equation}
f_z=nf_d + \frac{1}{N\Delta t}\frac{R_{n+1}}{R_{n+1} + 1} + \mathcal{O}
\left(10^{-3}\right), 
\end{equation}
\begin{equation}
f_z=nf_d + \frac{1}{N\Delta t}\frac{R_{n-1}}{R_{n-1} - 1}
+ \mathcal{O} \left(10^{-3}\right).
\end{equation}
Averaging these two formulas gives:
\begin{equation}
  \label{eqn:omeg_interp}
  f_z=nf_d +\frac{1}{2N\Delta t} \left[ \frac{R_{n+1}}{1 + R_{n+1}}
  + \frac{R_{n-1}}{R_{n-1} - 1}\right] +\mathcal{O} \left(10^{-3}\right)
\end{equation}

\begin{figure}
  \includegraphics[width=8.5cm]{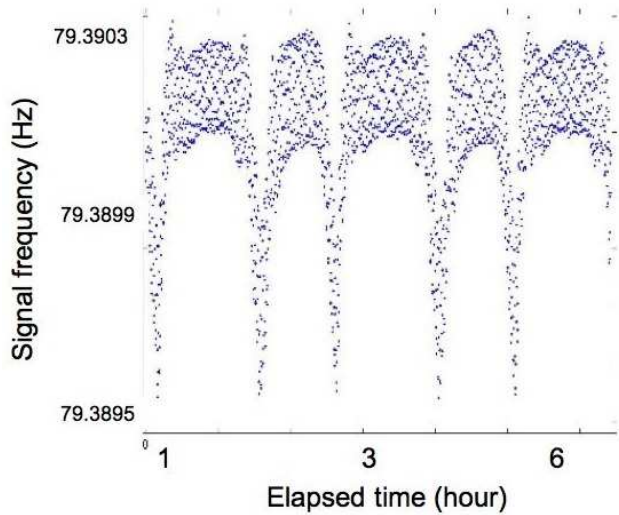}
  \caption{\label{fig:interp_Output}(Color online)
  Frequency $f_z$ from interpolation, gyroscope 1, Aug. 19 2004 }
\end{figure}

Therefore, the frequency interpolation yields a determination of the HF
signal's frequency $f_z$, whose error is
$\mathcal{O}(x_n/N) \lesssim \pi/N$, or $10^{-3}$ Hz.
Indeed, the frequency computed by this method typically showed a
$100$ $\mu$Hz spread, as shown on figure \ref{fig:interp_Output}.

The result derived in (\ref{eqn:omeg_interp}) was obtained for the simple
case of a signal with only one harmonic.
However, the HF signal contains many harmonics of its main frequency,
and as explained in \cite{salo,koza}, the odd harmonics have a significantly
larger amplitude. The interpolation procedure is therefore separately
applied to the fundamental frequency, $f_z$,
as well as to the $3^{rd}$ and $5^{th}$ harmonics of the FFT data.
The results from each of those 3 interpolations are then averaged in order
to obtain a determination of the signal frequency $f_z$.
Results for a 6-hour long data stretch for gyroscope 1 are shown
in figure \ref{fig:interp_Output}.

The formula (\ref{eqn:omeg_interp}) for the signal's frequency is valid
when no window is applied.
Applying windows to a signal consists of multiplying it by a carefully
chosen function -for example, a sine square- so the frequency components
away from the main frequency are attenuated \cite{harr,window}. 
In the case of the windowed signal, the analytical expression of $f_z$
is more complex but the principle of the derivation remains exactly the same.

\section {Microhertz level signal frequency determination through
phase differencing}
\label{part:micro}

The interpolation method relies on the use of the amplitude of each
independent FFT and, in practice, yields a $1$ mHz or better frequency
resolution. We now present an additional method called phase differencing
that uses the phase of the peak FFT bin, which is the one closest to the
frequency of interest, at two different times.
This method yields more than two additional orders of magnitude
in frequency resolution.
Phase differencing consists of using a precise measurement of the change
in the phase of the FFT computed on two distinct 1.86 second-long data
stretches. Knowing the time elapsed between the two data stretches,
we can determine the signal frequency. However, to resolve
the $2\pi$ ambiguity, an initial estimate of this frequency is necessary.
Indeed, let $\phi_1$ be the phase of the FFT obtained at time $t_1$
and $\phi_2$ the phase of the FFT obtained at time $t_2$.
We can assume that the frequency $f_z$ of the signal is constant if
$t_1$ and $t_2$ are close (we indeed show later that the characteristic
time of the decay in $f_z$ is 7,000 - 25,700 years, depending on the gyroscope).
We then have:

\begin{equation}
  \label{eqn:f_est}
  \phi_2-\phi_1=2\pi f_z(t_2-t_1)-2\pi K,
\end{equation}
where $K$ is an integer such that $\phi_2-\phi_1$ is between 0 and 2$\pi$.
We then need to know the value of $K$ in order to extract the frequency
$f_z$ from the previous equation.
This requires knowing an estimate $f^{est}_z$ of the signal frequency so that: 
\begin{equation}
  \label{eqn:f_est_cond}
  2\pi f^{est}_z(t_2-t_1)-2\pi K \in \left[ 0,2\pi \right].
\end{equation}
The value of $K$ then has to be chosen to ensure condition
(\ref{eqn:f_est_cond}). In order for $K$ to be known,
$f^{est}_z$ must lie within $1/10$ Hz of the real value $f$ since
$(t_2  - t_1)=10$ sec.
This resolution is easily achieved by the interpolation algorithm.
Furthermore, this $1/10$ Hz requirement is more stringent than the
$0.54$ Hz precision of the FFT:
the result of the interpolation is thus needed to carry out the
phase differencing.

\begin{figure}
  \begin{center}
    \includegraphics[width=8.5cm]{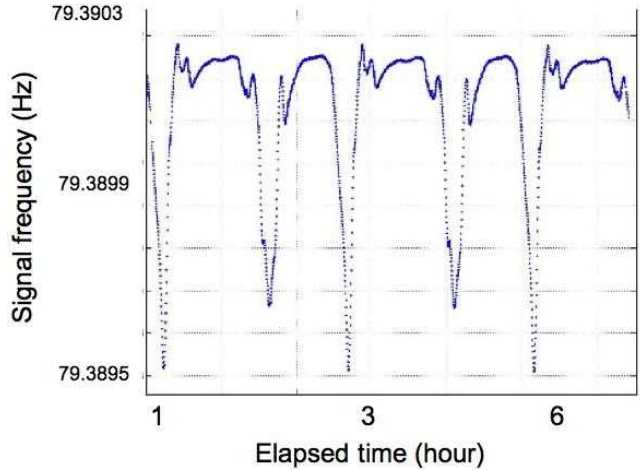}
  \end{center}
  \caption{\label{fig:phs_diff_output}(Color online)
  Frequency $f_z$ from phase differencing, gyroscope 1, Aug. 19 2004 }
\end{figure}

Then, once $K$ is computed, equation (\ref{eqn:f_est}) can trivially
be solved for the signal frequency $f_z$.
This new estimate has a very good precision. Indeed,
the phase is known with a conservative error of 10 arcsec
($5 \times 10^{-5}$ rad),
which at the gyroscope spin speed corresponds to a timing error of
approximately $7 \times 10^{-8}$ sec,
which may be compared to the short-term on-board clock accuracy
of approximately $10^{-8}$ sec.
The main error term is thus the phase measurement,
and for a  $(t_2-t_1)=10$ sec interval,
we obtain a precision better than 5 $\mu$Hz.
Figure \ref{fig:phs_diff_output} shows the extra resolution
that can be achieved with phase differencing.
It should be compared with figure \ref{fig:interp_Output},
which only shows the output of the interpolation on the same data stretch.

The large modulations in $f_z$ are at polhode frequency and are a
systematic error in the estimation method.
Indeed, we used a simple model for the
HF signal $z_{\text{HF}}$ which only takes into account the harmonics of $f_z$,
whereas harmonics of the polhode frequency also contribute to the
signal's spectrum. These frequencies then leak into our current
measurement of $f_z$.
The model introduced in the last section of this article takes into
account all the frequency components of the HF signal,
and yields an estimate of $f_z$ free of large variations at polhode frequency.

As a conclusion, by interpolating each FFT and using phase
differencing between FFTs, we can obtain an estimate of the signal
frequency with an error of a few $\mu$Hz.
We have thus gained 5 orders of magnitude as compared to the FFT bin
width of $0.54$ Hz.
Furthermore, this determination only relies on estimates of the FFT
for a small number of frequencies being transmitted from the spacecraft.
It then has the double advantage of requiring little bandwidth and
using less on-board CPU time, since the interpolation and the phase differencing
are carried out on the ground. 
Typical values for $f_z$ obtained after phase differencing are given in
table \ref{tbl:spin_speed}.
They have been measured on Feb. $6^{th}$ 2005 at 07:39 GMT.

\begin{table}
  \begin{ruledtabular}
    \begin{tabular}{cc}
      gyro& signal frequency $f_z$ (Hz) \\
      \hline
      1 & 79.38715\\
      2 & 61.81759\\
      3 & 82.09202\\
      4 & 64.85030\\
    \end{tabular}
  \end{ruledtabular}
  \caption{Typical values for signal frequency $f_z$ obtained from
  phase differencing}
  \label{tbl:spin_speed}
\end{table}

\begin{figure}
  \begin{center}
    \includegraphics[width=8.5cm]{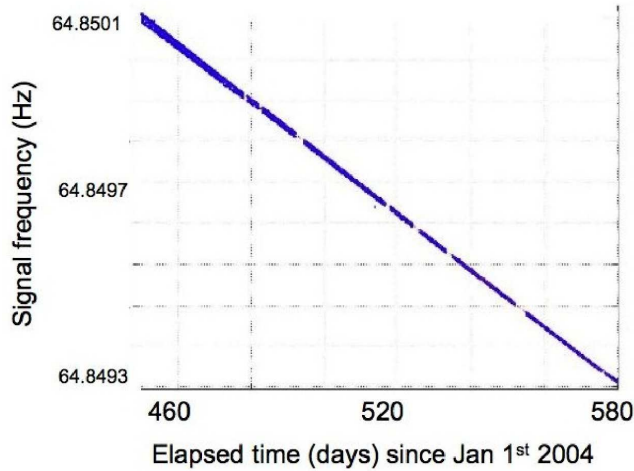}
  \end{center}
  \caption{\label{fig:sdown}(Color online)
    HF signal frequency $f_z$ on gyroscope 4 over 4 months }
\end{figure}

Figure \ref{fig:sdown} shows a linear decay of the frequency $f_z$
for gyroscope 4 over four months.
This decay was observed on all four gyroscopes.
As the polhode frequency $f_p$ increased slowly throughout the experiment,
equation (\ref{eqn:fz}) implies that the decay in frequency can be
attributed to a decay in $\dot{\phi}_s$.

The FFT is taken on-board every $10$ seconds for approximately $40\%$
of each orbit, a period called Guide Star Invalid (GSI).
During the rest of the orbit, the Guide Star Valid (GSV) period,
the on-board FFT algorithm is turned off in order to allow the CPU capacity
to be used to track the guide star:
we then have about $4000$ frequency measurements per day with a 10-second
spacing, each one with a $5$ $\mu$Hz standard deviation.
A covariance analysis yields a $0.1$ nHz/sec uncertainty
on the value of the signal frequency decay rate using 24 hours worth of data
(assuming 16 GSI periods per day,
each lasting 40 minutes and containing 240 samples,
and that for each GSI period a frequency offset must be estimated
together with the decay rate). 
An averaging time of 24 hours is chosen because there are
sometimes large gaps in the FFT data,
ranging from a few hours to a few days,
that occurred at roughly 1-2 day intervals that preclude
longer averaging times.
These gaps are related to details of the spacecraft operations and,
on rare occasions, to spacecraft anomalies.

Typical values for the frequency decay rate and the characteristic time
of the decay are given in table \ref{tbl:spin_down}.
\begin{table}
  \begin{ruledtabular}
    \begin{tabular}{ccc}
      gyro & frequency decay rate & characteristic time\\
        & (nHz/sec) &  of decay (years) \\
      \hline
      1 & 0.16 & 15,800  \\
      2 & 0.14 & 13,400  \\
      3 & 0.36 & 7,000   \\
      4 & 0.08 & 25,700 \\
    \end{tabular}
  \end{ruledtabular}
  \caption{Typical decay rate and characteristic time for the frequency
    $f_z$ of the HF signal}
  \label{tbl:spin_down}
\end{table}

\section{Verification using snapshot data}

The results obtained so far are derived from the analysis of the FFT data.
To increase the level of confidence in these results,
they have been checked by an independent verification using the snapshot data.
These data are more voluminous, hence much richer, than the FFT data:
a snapshot comprises 4096 points, whereas an FFT compacts it into 19 points.
Typically, snapshots are transmitted about every 40 seconds during
approximately an hour, separated by gaps ranging from an hour to up
to two days. 

A two-step algorithm similar to the procedure followed in the FFT analysis
was implemented in order to determine the signal frequency from the
snapshot data. First, a Fast Fourier Transform of each snapshot was
performed and interpolated in order to produce an initial estimate
of the spin frequency.
A two-point interpolation algorithm was used (instead of three-point)
to simplify the coding. 
A least-squares fit of each snapshot to a sum of up to $51$ harmonics
of the frequency obtained from the interpolation was then performed.

Secondly, a nonlinear fit to the snapshot data for
the signal frequency, $f_z$, and the
amplitudes of the sine and cosine components of $2\pi f_z \, t$
and its harmonics
(up to the 51st) was performed
using the \textit{NonLinearRegress} routine available in Mathematica.
The initial conditions were, for $f_z$, the value obtained after the
interpolation and, for the harmonics coefficients,
the output of the linear fit. 

This two-step analysis yielded an independent estimation of the signal
frequency as well as a large number of harmonics coefficients using
a data set and a method which was different from those used in the FFT analysis.

\begin{figure}
  \begin{center}
    \includegraphics[width=8.5cm]{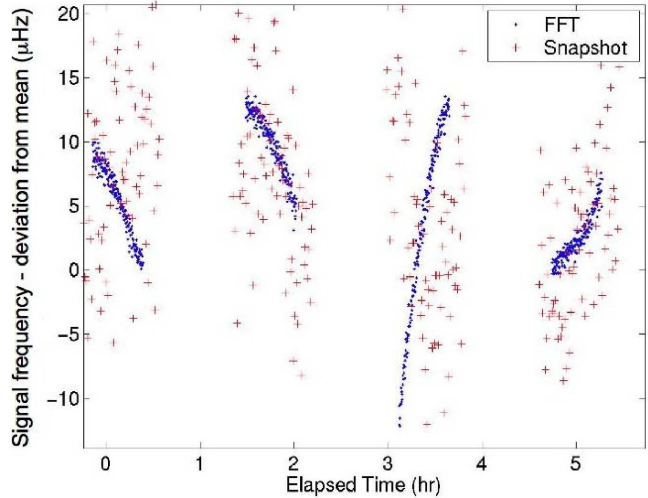}
  \end{center}
  \caption{\label{fig:spin_speed_comp}(Color online)
    $f_z$ obtained from FFT and snapshots, gyroscope 4, Feb. 7 2005 } 
\end{figure}

Figure \ref{fig:spin_speed_comp} compares $f_z$ determined from the FFT
and snapshot data on gyroscope 4 for February $7^{th}$ 2005.
Note that, like for the FFT data, the estimation of $f_z$ obtained
from the snapshot data shows variations at the polhode frequency.
This is expected as the model for the signal used in this determination
also neglects the harmonics of polhode. 
These results are typical and have been observed on all the
gyroscopes throughout the mission.
The signal's frequency obtained from the snapshots is typically noisier than
the one obtained from the FFT (the spread is on the order of $30$ $\mu$Hz).
The higher level of noise is likely due to the on-board
analog-to-digital converter, which exhibited noise at all multiples of 10 Hz.
Nevertheless, both time histories coincide strikingly despite the difference
in data sets and in algorithms, which increases our confidence
in the measurement.

Since more information is contained in the snapshot data, including harmonics
higher than the $\mathrm{5^{th}}$,
it was used as a reference for the remainder of the analyses presented in
this paper.
The analog-to-digital converter noise is later estimated and subtracted
from the snapshot data to remove its effects.
A more detailed discussion of the
justification for this choice can be found in \cite{salo}.
The last step of the method indeed dramatically increases the signal's
frequency resolution, which fully offsets the poorer quality of the
determination obtained from the snapshots.

\section{Nanohertz level signal frequency determination}

\subsection{Overview}

The interpolation and phase differencing methods allowed a determination
of the signal frequency $f_z$ with a microhertz level accuracy.
As the error in the determination of $f_p$ is 10 nHz,
formula (\ref{eqn:fz}) indicates that a better resolution in $f_z$
can be achieved if $\dot{\phi}_s$ is measured accurately.

In section \ref{part:micro} of this paper, it is shown that the linear
decay in the measured frequency $f_z$ is due to a decay in $\dot{\phi}_s$.
The expression given in (\ref{eqn:first_dotphis}) for $\dot{\phi}_s$ shows
that the decaying term is necessarily the angular momentum $L$.
We can then write:
\begin{equation}
  \label{eqn:phisdot}
    \dot{\phi}_s=2\pi(C_1 +C_2(t-t_0))\left[ 1 + \frac{I_3-I_1}{I_1}
      \frac{1}{1+\alpha^2 sn^2(\tau,k^2)}\right],
\end{equation}
where $t_0$ is a chosen time origin.

An accurate determination of the coefficients $C_1$ and $C_2$ is thus
necessary to estimate $\dot{\phi}_s$.
This last section describes how this estimation was carried out.
The underlying principle is to expand the HF SQUID signal in harmonics of
$\phi_s$. A model for the complex amplitude $H_n(t)$ of the $n^{th}$
harmonic of $\phi_s$ has been derived in \cite{firstTFM} and is given
in its compact form in equation (\ref{eqn:lin_model}):
the values $\hat{C}_1$ and $\hat{C}_2$ yielding the time history for
$\phi_s (t)$ such that the measured harmonics $H_n(t)$
are best accounted for by the
model are taken as the best estimates of $C_1$ and $C_2$.

Once $C_1$ and $C_2$ are known, the value of $\dot{\phi}_s$ can be computed
from (\ref{eqn:phisdot}), and the signal's frequency $f_z$ can be obtained
from (\ref{eqn:fz}).

\subsection{A cost function to estimate $\phi_s$}

Reference \cite{firstTFM} gives an expansion of the HF SQUID signal
$z_{\text{HF}}$ in harmonics of $\phi_s$:
\begin{equation}
  \label{eqn:HF_sig}
  z_{\text{HF}}(t)= \sum_{n=-\infty , n\neq0}^{\infty}H_{n}(t)e^{-in(\phi_s(t))}, 
\end{equation}

A model for $H_n(t)$ with $n$ odd can be derived using a similar procedure
to that used in \cite{firstTFM}, and this has been done in \cite{koza, salo}.
This model states that $H_n(t)$ depends on time only through the angles
$\phi_p(t)$ and $\gamma(t)$, determined in \cite{salo}, so that:
$$
H_n(t)=H_n(\phi_p(t),\gamma(t)).
$$
This model is furthermore linear in the coefficients of the expansion of
the magnetic potential in the body-fixed frame.
In other words, $H_n(\phi_p(t),\gamma(t))$ is linear in a state vector
$\vec{A}$ whose coefficients are constant as they only depend on the
body-fixed magnetic potential distribution.
We call $M_n(\phi_p(t),\gamma(t))$ the observability matrix for this linear
model. Its expression is given in \cite{salo}. We can thus write:
\begin{equation}
  \label{eqn:lin_model}
    H_n(\phi_p(t),\gamma(t))=M_n(\phi_p(t),\gamma(t))\vec{A}.
\end{equation}
Therefore, if the Euler angle $\phi_s(t)$ is known accurately,
$H_n(t)$ is known from (\ref{eqn:HF_sig}), and by running a linear least
squares algorithm, we can estimate the vector $\vec{A}$ that minimizes
the residuals $J_1$ where:
\begin{equation}
  \label{eqn:min_resid}
     J_1=\min_{\vec{A}}\Big|\Big|H_n(t) - M_n(\gamma(t),\phi_p(t))
      \vec{A}\Big|\Big|.
\end{equation}

However, errors in the estimation of $C_1$ and $C_2$ lead to an approximate 
value $\phi_s^{app}(t)$ for the third Euler angle.
The coefficient $H_n^{app}(t)$ obtained from the expansion (\ref{eqn:HF_sig})
of the HF signal in harmonics of $\phi_s^{app}(t)$ is therefore also
approximate. Consequently, the model (\ref{eqn:lin_model}) does not
apply exactly: the vector $\vec{A}^{app}$ that solves the
least squares problem (\ref{eqn:lin_model}) with the approximate value
$H_n^{app}(t)$ for $H_n(t)$ yields a minimum norm $J_1^{app}$ of the
residuals which is larger than the value $J_1$ defined in (\ref{eqn:min_resid}).

Therefore, the best estimate $\hat{\phi}_s$ of the third Euler angle is the
one that yields the smallest value for the residual $J_1^{app}$.
In other words, $\hat{\phi}_s$ minimizes the functional $J$ defined by
\begin{equation}
  \label{eqn:functional}
    J[\phi_s]=\min_{\vec{A}}\Big|\Big|H_n[\phi_s](t) - M_n(t)\vec{A}\Big|\Big|.
\end{equation}
The notation $J[\phi_s]$ means that the argument of the functional
J is the function $\phi_s$.

In reference \cite{salo}, an expression for $\phi_s$ is given based on
the integration of the expression (\ref{eqn:phisdot}) for $\dot{\phi}_s$:
\begin{eqnarray}
  \phi_s(t)&&=\phi_{s_{0}} + 2\pi C_1 \left[(t-t_0)
    + \Pi_1(t,t_0,I_1,I_2,I_3,f_p)\right] 
    \nonumber\\	
    && - 2\pi C_2 \left[\frac{1}{2} (t-t_0)^2+ \Pi_2(t,t_0,I_1,I_2,I_3,f_p)
    \right].
  \label{eqn:spin_phase}
\end{eqnarray}

In this expression, $\phi_{s_0}$ is $\phi_s$ at the time origin $t_{0}$.
The terms $\Pi_1$ and $\Pi_2$ are integrals of the Jacobi elliptic
function $sn$ and depend only on the moments of inertia and the polhode
frequency.
The function $\phi_s$ is thus fully determined by three variables:
$\phi_{s_0}$, $C_1$ and $C_2$.

Using equation (\ref{eqn:spin_phase}), we see however that the initial
value $\phi_{s_0}$ is just a constant shift to the angle $\phi_s(t)$.
From the expansion (\ref{eqn:HF_sig}),
a constant shift in $\phi_s(t)$ yields a constant phase shift
$e^{in\phi_{s_0}}$ in $H_n[\phi_s]$.
In the expression (\ref{eqn:functional}),
we can factor out this constant phase shift,
so that the term $\vec{A}$ becomes $\vec{A}e^{-in\phi_{s_0}}$.
And since $J[\phi_s]$ is a minimum over all values of $\vec{A}$,
its value is then independent of $\phi_{s_0}$.
Consequently, the best estimates of $C_1$ and $C_2$ minimize $J[\phi_s]$.

In other words, the estimation of $C_1$ and $C_2$ is a minimization problem
of a function $J_2$ of two variables:
\begin{equation}
  \label{eqn:min_problem}
    J_2(C_1,C_2)=\min_{\vec{A}}\Big|\Big|H_n(C_1,C_2,t)
      - M_n(\gamma(t),\phi_p(t))\vec{A}\Big|\Big|.
\end{equation}

We have thus constructed a cost function $J_2$ which can be minimized in
order to estimate the two coefficients $C_1$ and $C_2$. 

This frequency determination method is general.
Indeed, it only supposes that a model for the harmonics of the signal's
frequency is available: the frequency yielding the harmonics best fit by the
model is then the best estimate of the signal's frequency. 

Note that in the specific case of the GP-B gyroscopes,
the modeled quantities are not exactly the harmonics of the signal's
frequency $f_z$ of the signal but rather of $\dot{\phi}_s=2\pi(f_z-f_p)$,
where $f_p$ is known accurately.

\subsection{Implementation}

We now introduce the algorithm developed by the authors to estimate
$C_1$ and $C_2$.

A nonlinear minimization routine based on a modification of the
Nelder-Mead (NM) simplex \cite{avri}, used in Matlab's \emph{fnimsearch.m}
function, was implemented.
A simplex is a set of $n + 1$ points in an $n$-dimensional space
such that the points are not degenerate.
In 2-space the simplex is a triangle.
A detailed account of how this algorithm works can be found in \cite{koza}. 
This nonlinear simplex routine was chosen because of its
high accuracy for non-smooth objective functions compared with gradient
or Gauss-Newton methods, implemented in Matlab's $lsqnonlin.m$ function
for example.
The Nelder Mead simplex is a nonlinear minimization algorithm and as
such requires the input of initial conditions.
Since we know \emph{a priori} the region of possible values for $C_1$ and
$C_2$, we utilize a NM simplex method that is modified to include
a constraint on the search domain.
From equation (\ref{eqn:phisdot}), we find that, for the GP-B gyroscopes: 
$$
\dot{\phi}_s(t_0)=2\pi C_1(1+\mathcal{O}\left(10^{-6}\right))
$$
And since:
$$\dot{\phi}_s=2\pi(f_z-f_p)$$
we find:
$$
C_1=(f_z(t_0)-f_p(t_0))(1+\mathcal{O}\left(10^{-6}\right))
$$
Using the measurement of $f_z$ obtained previously,
we thus have an initial condition for $C_1$ as well as a search interval
whose width is on the order of $100$ $\mu$Hz.

Similarly, we use the measurement of the decay in $f_z$ to find an
initial condition and a search interval for $C_2$.
As $|df_p/dt|$ can be as large as $1/10~df_z/dt$, the relative size
of the search interval is larger than for the coefficient $C_1$:
we opted for a search interval whose size is 25\% of the measured decay
rate of $f_z$. 

Using these two initial conditions and search intervals,
we define a search region in the \Iddd plane.
The NM simplex algorithm will then search for the values of
$C_1$ and $C_2$ within this region that minimize the cost function $J_2$.

\begin{figure}
  \begin{center}
    \includegraphics[width=8.5cm]{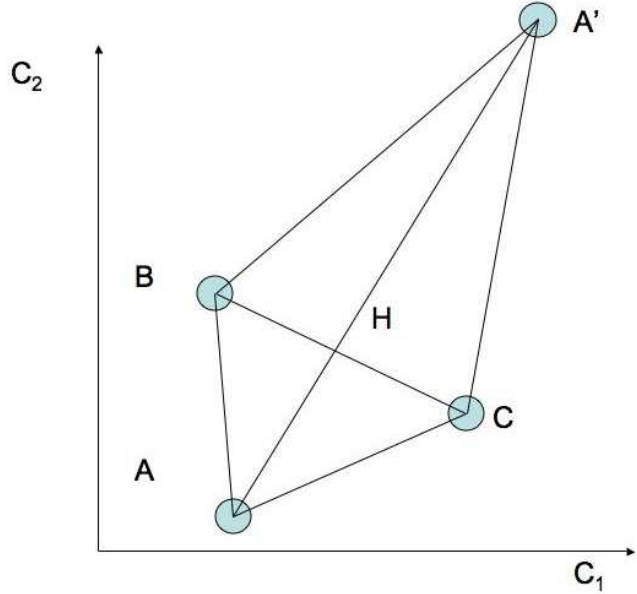}
  \end{center}
  \caption{\label{fig:NM_expand}(Color online)
    Expansion of the simplex}
\end{figure}

\begin{figure}
  \begin{center}
    \includegraphics[width=8.5cm]{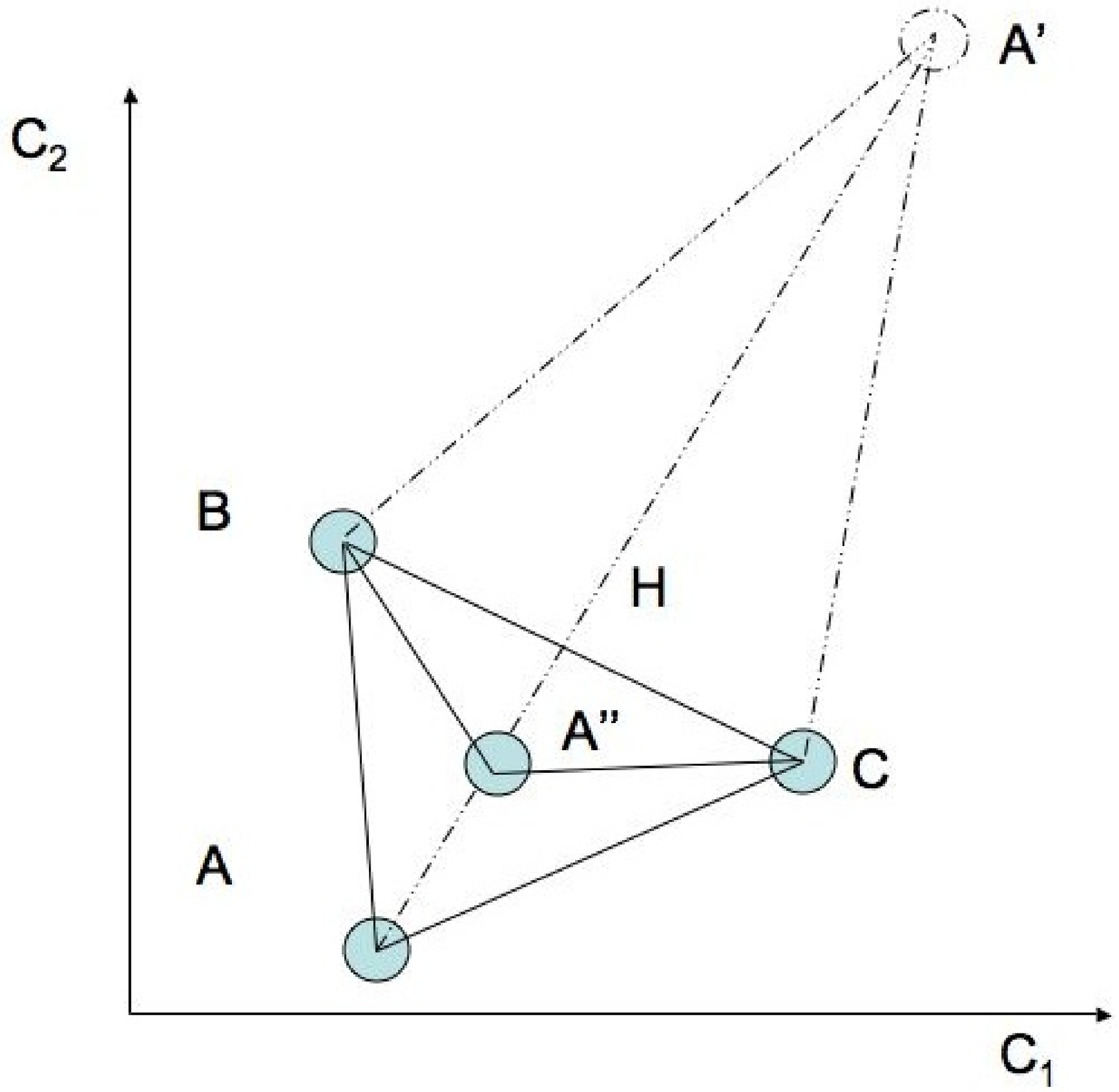}
  \end{center}
  \caption{\label{fig:NM_compac}(Color online)
    Simplex honing on a minimum : case  $J_2(A')>J_2(A)$}
\end{figure}

The NM algorithm is implemented in Matlab's built-in \textit{fminsearch.m}
routine.
The algorithm first evaluates the cost function $J_2$ on a set of three
points $A$, $B$ and $C$, called a simplex, within the search region.
The Matlab algorithm finds on which vertex $A$ of the simplex the function
has the maximum value. Let $B$ and $C$ be the other two vertices of the simplex
and $H$ the center of segment $BC$.
The algorithm evaluates the function at a point $A'$ such that $AA'$
is orthogonal to $BC$ and $AA'=\delta AH$ with $\delta=3$.
(see figure \ref{fig:NM_expand}).
This procedure is called an 'expansion'.
Indeed, the simplex $A'BC$ has a
bigger area than $ABC$.
If the cost function $J_2$ has a lower value in $A'$ than in $A$,
the algorithm is searching in the right direction in the \Iddd plane,
and the procedure repeats, now starting from $A'$.
If the function has a higher value in $A'$ than in $A$,
the expansion did not decrease the value of the function and
should then not be pursued.
The cost function is then evaluated at a new point $A''$ inside the
$ABC$ simplex  (see figure \ref{fig:NM_compac}),
such that $AH = \rho HA''$, where $\rho = 2$,
and the procedure repeats.
Matlab's algorithm stops when the dimension of the simplex,
that is the length of its longest side, is lower than the specified tolerance.

There are two main limitations in the available Matlab routine.
First, only one tolerance can be specified.
This is in our case a real issue as both the absolute and relative
tolerance on \Id and \Idd differ by many orders of magnitude.
Furthermore, the algorithm does not provide any way to decrease the
likelihood of honing in on a local minimum.

Matlab's \emph{fminsearch.m}
routine was thus modified to address these two issues.
First, a simple but crucial change was brought to \textit{fminsearch.m}
so that the termination tolerances on the two dimensions were allowed to
differ: the modified NM algorithm terminates only when the length of the
simplex along both the \Id and \Idd dimensions are smaller than values
yielding the required precision in $\dot{\phi}_s$:
these numbers are referred to as "termination tolerances".

After examining several algorithms for decreasing the
likelihood of reaching a local minimum,
the following steps were determined to be the most effective for
finding the global minimum, when applying the above model to the
snapshot data.
The first one is to add an extra expansion when the original algorithm
terminates: in case a value $J_2$ lower than the current minimum is found,
the algorithm restarts. Once a minimum is attained, a $20\times20$ grid in the
\Iddd plane is then created,
each point being spaced by the termination tolerance.
$J_2$ is evaluated at each of these points.
If one of the 400 points in this local grid yields a lower value of $J_2$
than the current minimum, the algorithm is restarted at this point.
The algorithm then terminates and returns the values of \Id and \Idd
that minimize $J_2$.
As a third and last step, we then restart this algorithm twice such that
at each iteration we use the same initial search region and the current
best guess of \Id and \Idd as initial conditions.
The new best estimates of \Id and \Idd are then compared with the
previous ones and it has been observed that 3 iterations are sufficient
to obtain estimates within a few times the termination tolerances.

The estimation of \Id and \Idd was performed using the modified NM algorithm
to minimize the cost function $J_2$.
This cost function is however based on a spherical harmonics model
for the fluxon distribution around the gyroscope,
and the order of this expansion, conventionally denoted by the letter $l$,
is in theory a free parameter.
Nevertheless, early estimations and fits to the SQUID data suggested a range
of values for $l$.
The NM search was thus repeated for all reasonable values of $l$.
The optimal values for \Id and \Idd were obtained by averaging the estimates
corresponding to different values of $l$.

The standard deviation of the ensemble of estimates of $C_1$ and $C_2$ was on
the order of $10$ nHz for $C_1$ and $10^{-4}$ nHz/sec for $C_2$.
These results were obtained using one-day batches.
Therefore, a conservative estimate of the
uncertainty in $\dot{\phi}_s$ at the end of a one-day batch is:
$10$ nHz + $10^{-4}$ nHz/sec $\times$ 86,400 sec $\approx$ $20$ nHz
(see Eq. (\ref{eqn:phisdot})).
The uncertainty in the signal frequency, $f_z = \dot{\phi}_s/2\pi + f_p$,
is then conservatively estimated to be $30$ nHz, for a $10$~nHz uncertainty
in the polhode frequency, $f_p$.
This uncertainty bounds the uncertainties for all gyroscopes during the
entire science mission.
For a gyroscope spinning at 60-80 Hz,
this represents a relative error of $5\times10^{-10}$.
Unlike the previous two approaches,
the polhode motion of the gyroscopes is modeled explicitly.
Therefore, the polhode frequency harmonics are completely separated from the
gyroscope spin frequency estimates.
The value of the signal frequency and decay rate for February $6^{th}$ 2005
are given in tables \ref{tbl:spin_speed2} and \ref{tbl:spin_down2}.
They can be compared with the results obtained with the phase differencing
method on the same day in tables \ref{tbl:spin_speed} and \ref{tbl:spin_down}.

\begin{table}
  \begin{ruledtabular}	
    \begin{tabular}{cc}
      gyro& signal frequency (Hz) \\
      \hline
      1 & 79.38746144\\
      2 & 61.81765160\\
      3 & 82.09121932\\
      4 & 64.85036174\\
    \end{tabular}
  \end{ruledtabular}
  \caption{Signal frequency on February $6^{th}$ 2005 at 07:39:00 GMT}
  \label{tbl:spin_speed2}
\end{table}

\begin{table}
  \begin{ruledtabular}	
    \begin{tabular}{cc}
      gyro& frequency decay rate (nHz/sec) \\
      \hline
      1 & 0.1581 \\
      2 & 0.1432 \\
      3 & 0.3634\\
      4 & 0.0803\\
    \end{tabular}
  \end{ruledtabular}
  \caption{Signal frequency $f_z$ decay rate on February $6^{th}$ 2005}
  \label{tbl:spin_down2}
\end{table}

\section{Conclusion}

This paper describes three successive algorithms for estimating the frequency
of a digital signal.
The first two of them, interpolated FFT and phase differencing,
known previously, were properly adjusted for the needs of the analysis
of the HF signal obtained in the Gravity Probe B experiment.
The third method is new;
it is based on the accuracy achieved by the previous two approaches,
and involves a nonlinear estimation of a slowly changing frequency
when many harmonics of another time-varying frequency are present.
The last method requires accurately modeling the rigid body motion
of a nearly torque-free gyroscope, and utilizes the specially modified
Nelder – Mead simplex method.

These algorithms are applied to a particular problem of precisely
determining the spin frequency of the ultra-precise GP-B gyroscopes.
The measured HF SQUID signal was used for this analysis
whose signal to noise ratio was $\sim 10^5$.
We achieved the ultimate relative accuracy of 5 parts in $10^{10}$,
corresponding to the 30 nHz absolute accuracy for a gyro spinning at 60-80 Hz.
This accuracy level was necessary for the estimation of trapped magnetic
flux on the surface of each gyroscope rotor to 1\%,
the resulting time variations in the gyro readout scale factor to $10^{-4}$,
and ultimately the relativistic gyro drift rate to 20 marcsec/yr.


\end{document}